\documentstyle [12pt] {article}

\begin{document}

\title{The average distance of the $n$-th neighbour in a uniform distribution
of random points}

\author{P. Bhattacharyya, B. K. Chakrabarti and A. Chakraborti \\~\\
 {\small Saha Institute of Nuclear Physics,}\\
 {\small Sector - I, Block - AF, Bidhannagar, Kolkata 700 064, India}}

\maketitle

\begin{abstract}

We first review the derivation of the exact expression for the
average distance $\left \langle r_n \right \rangle$ of the $n$-th
neighbour of a reference point among a set of $N$ random points
distributed uniformly in a unit volume of a $D$-dimensional
geometric space. Next we propose a \lq mean-field\rq~theory of
$\left \langle r_n \right \rangle$ and compare it with the exact
result. The result of the \lq mean-field\rq~theory is found to
agree with the exact expression only in the limit $D \to \infty$
and $n \to \infty$. Thus the \lq mean-field\rq~approximation is
useless in this context.

\end{abstract}

\newpage

\section {Introduction to the average $n$-th neighbour distance}

\indent Consider $N$ (a large number) points distributed randomly
and uniformly in a unit volume of a $D$-dimensional geometric space.
A point is said to be the $n$-th neighbour of another (the reference point)
if there are exactly $n - 1$ other points that are closer to the latter
than the former.
The average distance to the first neighbour is exactly known
\cite{Chandrasekhar1943}; though originally calculated in three
dimensions the method can be used for any finite dimension $D$~:
The probability distribution $P(r_1) {\mathrm d}r_1$ of the
first neighbour distance is defined by the probability of
finding the first neighbour of a given reference point at a distance
between $r_1$ and $r_1 + {\mathrm d}r_1$~:

\begin{equation}
P(r_1) \: {\mathrm d}r_1 = \left [ 1 - V(r_1) \right ]^{N - 1} \:
 (N - 1) \: {\mathrm d}V(r_1) ,
 \label{eq:prob-1}
\end{equation}

\noindent where $V(r_1) = \pi^{D/2} \cdot (r_1)^D / \Gamma(D/2 + 1)$
is the volume of a $D$-dimensional hypersphere of radius $r_1$
centered at the reference point. The average first neighbour distance
is defined as~:

\begin{equation}
\left \langle r_1 \right \rangle = \int_0^R \: r_1 \: P(r_1) \:
 {\mathrm d}r_1 ,    \label{eq:defn-av1}
\end{equation}

\noindent where $R$ is the radius of a $D$-dimensional hypersphere
of unit volume~:

\begin{equation}
R = {1 \over \pi^{1/2}} \: \left [\Gamma \left ({D \over 2}
     + 1\right )\right ] ^{1/D} .    \label{eq:unitvol-rad}
\end{equation}

\noindent With the probability distribution of equation \ref{eq:prob-1}
we get

\begin{eqnarray}
\left\langle r_1\right\rangle & = & \int_0^1  r_1 \: [1 - V(r_1)]^{N - 1} \:
 (N - 1) \: {\mathrm d}V(r_1)
 \nonumber \\
           ~ & = & {1 \over \pi^{1/2}} \:
 \left [ \Gamma\left({D \over 2} + 1\right ) \right ]^{1/D}
 \Gamma \left ( 1 + {1 \over D} \right ) \: \left ( 1 \over N \right )^{1/D} .
 \label{eq:av-1}
\end{eqnarray}

\indent Now we address the general problem~: What is the form of the
average $n$-th neighbour distance, for any finite $n$?
Though this is a problem of purely geometric nature,
the quantity $\left\langle r^{(D)}_N(n)\right\rangle$ is relevant
in physical and computational contexts; for example,
in astrophysics we need to know the average distance between
neighbouring stars distributed
independently in a homogeneous universe~\cite{Chandrasekhar1957},
and in the traveling salesman problem we need the average distance
of the neighbours of each site for estimating the optimal
path-length~\cite{Beardwood1959}.

\indent We proceed by extending the line of argument used in the
case of the first neighbour \cite{Chandrasekhar1943}
to the $n$-th neighbour. The probability
distribution of the $n$-th neighbour distance $r_n$ is defined as
the probablity $P(r_n) {\mathrm d}r_n$ of finding the $n$-th neighbour
of a given reference point at a distance between
$r_n$ and $r_n + {\mathrm d}r_n$.
This is a {\em conditional probability} because we look for the
$n$-th neighbour of a point when its first $(n - 1)$ neighbours
have already been located~:

\begin{equation}
P\left (r_n\right ) \: {\mathrm d}r_n= \left [ 1 - {V(r_n) -
 V(r_{n - 1}) \over
 1- V(r_{n - 1})}\right ]^{N - n} \:
 {(N - n) \: {\mathrm d}V(r_n) \over 1 - V(r_{n - 1})} .
 \label{eq:prob-n}
\end{equation}

\noindent The quantity $V(r_n)$ is the volume
of a $D$-dimensional hypersphere of radius $r_n$ centered at the
reference point. For a given reference point and its first
$n - 1$ neighbours the average $n$-th neighbour distance is
obtained as~:

\begin{equation}
\left \langle r_n\right\rangle_{\mathrm (particular)}
 = \int_{r_{n-1}}^R \: r_n \: P\left ( r_n\right) \: {\mathrm d}r_n
 \label{eq:particular}
\end{equation}

\noindent where, as before, $R$ is the radius of a $D$-dimensional
hypersphere of unit volume. The quantity 
$\left \langle r_n\right\rangle_{\mathrm (particular)}$
is a function of a particular $r_{n-1}$, $r_{n-2}$, $\ldots$, $r_1$
which are the distances of the first $n-1$ neighbours of the given
reference point. To calculate the ensemble average of $r_n$
the quantity $\left \langle r_n\right\rangle_{\mathrm (particular)}$ must be
averaged successively over the probability distributions of each of
the first $n-1$ neighbours~:

\begin{eqnarray}
\left \langle r_n\right\rangle &
 = & \int_0^R \: {\mathrm d}r_1 \: P(r_1) \:
   \int_{r_1}^R \: {\mathrm d}r_2 \: P(r_2) \: \cdots
   \int_{r_{n-3}}^R \: {\mathrm d}r_{n-2} \: P(r_{n-2})
   \nonumber \\
   ~ & ~ & \times
   \int_{r_{n-2}}^R \: {\mathrm d}r_{n-1} \: P(r_{n-1}) \:
   \int_{r_{n-1}}^R \: {\mathrm d}r_n \: r_n \: P( r_n)
 \label{eq:general}
\end{eqnarray}

\noindent where the probability distribution of the $i$-th neighbour
is given by equation \ref{eq:prob-n} with $i$ replacing $n$.
After a change in the order of the integrals in equation \ref{eq:general}~:

\newpage

\begin{eqnarray}
\left \langle r_n\right\rangle &
 = & (N - 1) (N - 2) \cdots (N - n) {\left [\Gamma\left ({D \over 2} 
                              + 1\right )\right ]^{1/D} \over \pi^{1/2}}
   \nonumber \\
   ~ & ~ & \times \int_0^1 \: {\mathrm d}V(r_n) \:
     \left [V(r_n)\right ]^{1/D} \:
              \left [1 - V(r_n)\right ]^{N - n}
   \int_0^{V(r_n)} \: {\mathrm d}r_1 \nonumber \\
   ~ & ~ & \times
   \int_{V(r_1)}^{V(r_n)} \: {\mathrm d}r_2 \:
   \cdots
   \int_{V(r_{n-3})}^{V(r_n)} \:
              {\mathrm d}r_{n-2} \:
   \int_{V(r_{n-2})}^{V(r_n)} \:
              {\mathrm d}r_{n-1}
 \label{eq:changed-general}
\end{eqnarray}

\noindent which gives the final form of the average $n$-th
neighbour distance~:

\begin{eqnarray}
\left \langle r_n\right \rangle & = &
 \int_0^1 \: \left ( \begin{array}{c}
                          N - 1\\
                          n - 1
                     \end{array}
             \right ) \: \left [V(r_n)\right ]^{n + (1/D) -1} \:
                      \left [1 - V(r_n)\right ]^{N - n} (N - n) \:
                      {\mathrm d}V(r_n) \nonumber \\
 ~ & = & {1 \over \pi^{1/2}} \:
       \left [\Gamma\left ({D \over 2} + 1\right )\right ]^{1/D} \:
       {\Gamma\left (n + {1 \over D}\right ) \over \Gamma(n)} \:
       \left ({1 \over N}\right )^{1/D} .
 \label{eq:percus-av}
\end{eqnarray}

\noindent This result was reported in \cite{Percus1996}.

\indent Next we consider fluctuations $\delta r_n$ occuring in
$r_n$. This can be calculated exactly for any neighbour $n$~;
the mean square deviation in $r_n$ from its average value is given by~:

\begin{eqnarray}
 \left (\delta r_n\right )^2 & = & \left\langle r_n^2\right\rangle
 - \left\langle r_n\right\rangle^2 
 \nonumber \\
 ~ & = & {1 \over \pi} \left [\Gamma\left ({D \over 2} + 1\right )\right ]
         ^{2/D}
         \left [{\Gamma\left (n + {2 \over D}\right ) \over \Gamma(n)}
              - {\Gamma^2\left (n + {1 \over D}\right ) \over \Gamma^2(n)}
         \right ] \left (1 \over N\right )^{2/D}
 \label{eq:fluct}
\end{eqnarray}

\noindent which vanishes as $D \to \infty$. This suggests that
the form of $\left\langle r_n\right\rangle$ for large $D$ can be
arrived at by neglecting fluctuations, an approach which
corresponds to mean-field theories in statistical mechanics.  

\section {A \lq mean-field\rq~theory}

\indent By the following \lq mean-field\rq~argument we derive an
expression for the average $n$-th neighbour distance in large
dimensions $D$. Since the average first neighbour distance can be
found easily, we derive $\left\langle r_n\right\rangle$
in terms of $\left\langle r_1\right\rangle$.
As before we consider $N$ (a large number of) random points distributed
uniformly within a unit volume of a $D$-dimensional geometric space.
We choose any one of them as the reference point and locate its
$n$-th neighbour. Neglecting fluctuations, which we can do for large
$D$, the distance between them is $r_n(N) \approx
\left\langle r_n(N)\right\rangle$. Keeping these two points
fixed we change the number of points in the unit volume to $N \alpha$
by adding or removing points at random; the factor $\alpha$ is arbitrary
to the extent that $N \alpha$ and $n \alpha$ are natural numbers. Since
the distribution of points is uniform, the hypersphere that had originally
enclosed just $n$ points will now contain $n \alpha$ points. Therefore,
what was originally the $n$-th neighbour of the reference point now
becomes the $n \alpha$-th neighbour. Since the two points under
consideration are fixed, so is the distance between them. Consequently,

\begin{equation}
\left\langle r_n(N)\right\rangle
 \approx \left\langle r_{n \alpha}(N \alpha)\right\rangle .
\end{equation}

\noindent Now we take $\alpha = 1/n$, so that

\begin{equation}
\left\langle r_n(N)\right\rangle
 \approx \left\langle r_1(N/n)\right\rangle ,
\end{equation}

\noindent which shows that the average $n$-th neighbour distance for
a set of $N$ random points distributed uniformly is approximately
given by the average distance for a depleted set of $N/n$ random
points in the same volume. Using the expression for
$\left\langle r_1(N)\right\rangle$ from equation \ref{eq:av-1} we get

\begin{equation}
\left\langle r_n\right\rangle \approx {1 \over \pi^{1/2}} \:
\left [ \Gamma\left ({D \over 2} + 1\right) \right ]^{1/D}
 \Gamma \left (1 + {1 \over D}\right ) \: \left ( n \over N \right )^{1/D} .
 \label{eq:mf-av-n}
\end{equation}

\noindent Since the above argument neglects fluctuations the result
of equation \ref{eq:mf-av-n} ought to be exact in the limit $D \to \infty$.

\indent The exact expression of $\left\langle r_n\right\rangle$
for a finite dimension $D$ is expected to reduce to the form of equation
\ref{eq:mf-av-n} as $D \to \infty$ where fluctuations do not affect.
For large $D$ equation \ref{eq:percus-av} takes the following form~:

\begin{equation}
\left\langle r_n\right\rangle \approx {1 \over \pi^{1/2}} \:
\left [ \Gamma\left ({D \over 2} + 1\right) \right ]^{1/D}
 \Gamma \left (1 + {1 \over D}\right ) \:
 \left (1 + {1 \over D} \sum_{k = 1}^{n - 1} {1 \over k} \right )
 \left ( 1 \over N \right )^{1/D} .
 \label{eq:percus-largeD}
\end{equation}

\noindent For the above expression to reduce to the form of equation
\ref{eq:mf-av-n} the sum $\sum_{k = 1}^{n - 1} {1 \over k}$
must be equal to $\log_e n$ which happens only in the limit
$n \to \infty$. Thus for any finite $n$ the exact result of equation
\ref{eq:percus-av} fails to produce the fluctuation-free form
of equation \ref{eq:mf-av-n} in large dimensions $D$. This shows that
the \lq mean-field\rq~approximation is useless in the present
context. However the \lq mean-field\rq~approach for
$\left\langle r_n\right\rangle$ may be used as a crude
approximation in other distributions (non-uniform) of random points
where an exact calculation is not possible beyond the first neighbour.

\section*{Acknowledgement}

\indent We thank D. Dhar, S. S. Manna and A. Percus for their comments.

\end{document}